\documentclass[11pt,preprint]{aastex}
\usepackage{times,color,wrapfigure,multicols}
\usepackage[hmargin=1in,vmargin=1in]{geometry}
\bibpunct{(}{)}{;}{a}{}{,}
\bibpunct{(}{)}{;}{a}{}{,}                        

\renewenvironment{thebibliography}[1]{%
  \begin{oldthebibliography}{#1}%
    \footnotesize 
    \setlength{\parskip}{0pt}                     
    \setlength{\itemindent}{-15pt}                
    \setlength{\itemsep}{0pt}                     
    \hyphenpenalty=9999
    \frenchspacing                                
}{\end{oldthebibliography}}

\let\oldbibliography\thebibliography
\renewcommand{\thebibliography}[1]{%
  \oldbibliography{#1}%
  \setlength{\itemsep}{0pt}%
}

\newcommand{\masyr}{mas~yr\ensuremath{^{-1}}}

\newcommand{\ha}{H\ensuremath{\alpha}}

\newcommand{\Teff}{T\ensuremath{_{\rm eff}}}

\shorttitle{Stellar Ages with \textit{Kepler}'s Two-Wheels}
\shortauthors{Dhital et al.}

\begin{document}

\title{A Gyrochronology and Microvariability Survey of the Milky Way's
  Older Stars Using \textit{Kepler}'s Two-Wheels Program}
\author{
  Saurav Dhital\altaffilmark{1, 7},
  Terry D. Oswalt\altaffilmark{1},
  Philip S. Muirhead\altaffilmark{2}, 
  Kolby L. Weisenburger\altaffilmark{2},
  Sydney A. Barnes\altaffilmark{3}, 
  Kenneth A. Janes\altaffilmark{2},
  Andrew A. West\altaffilmark{2},
  Kevin R. Covey\altaffilmark{4},
  S{\o}ren Meibom\altaffilmark{5}, 
  Trisha F. Mizusawa\altaffilmark{6}
}
\altaffiltext{1}{Embry-Riddle Aeronautical University, 600 South Clyde Morris Blvd., Daytona Beach, FL 32114, USA}
\altaffiltext{2}{Boston University, 725 Commonwealth Avenue, Boston, MA 02215, USA}
\altaffiltext{3}{Leibniz Institute for Astrophysics, Potsdam 14467, Germany}
\altaffiltext{4}{Lowell Observatory, 1400 West Mars Hill Road, Flagstaff, AZ 86001, USA}
\altaffiltext{5}{Harvard-Smithsonian Center for Astrophysics,  Cambridge, MA 02138, USA}
\altaffiltext{6}{Florida Institute of Technology, 150 W. University Blvd., Melbourne, FL, 32901, USA}
\altaffiltext{7}{Corresponding email: saurav.dhital@erau.edu}

\begin{abstract}
Even with the diminished precision possible with only two reaction
wheels, the \textit{Kepler} spacecraft can obtain mmag level,
time-resolved photometry of tens of thousands of sources. The presence
of such a rich, large data set could be transformative for stellar astronomy.
In this white paper, we discuss how rotation periods for a large
ensemble of single and binary main-sequence dwarfs can yield a
quantitative understanding of the evolution of stellar spin-down over
time. This will allow us to calibrate rotation-based ages
beyond $\sim$1~Gyr, which is the oldest benchmark that
exists today apart from the Sun. Measurement of rotation periods of M
dwarfs past the fully-convective boundary will enable extension of 
gyrochronology to the end of the stellar main-sequence, yielding 
precise ages ($\sigma \sim $10\%) for the vast majority of nearby
stars. It will also help set constraints on the angular
momentum evolution and magnetic field generation in these
stars. Our \textit{Kepler}-based study would be supported by a suite
of ongoing and future ground-based observations.
Finally, we briefly discuss two ancillary science cases, detection of
long-period low-mass eclipsing binaries and microvariability in white dwarfs and hot
subdwarf B stars that the \textit{Kepler} Two-Wheels Program would facilitate.
\end{abstract}

\section{Introduction}
\begin{wrapfigure}{R}{0.6\textwidth}
  \begin{center}
    \includegraphics[width=0.5\textwidth]{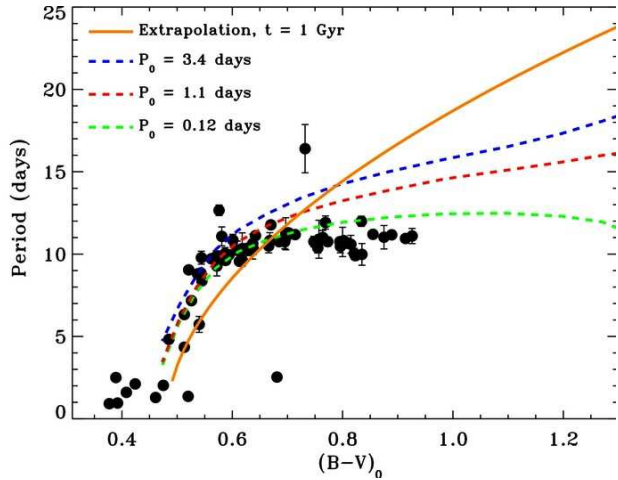}
    \vspace{-0.1in}
    \caption{\protect\small
      The rotation period--color diagram for the open cluster
      NGC~6811 as measured by \textit{Kepler} \citep{Meibom2011a}.
      The orange curve is an extrapolation of data from younger open
      clusters to 1~Gyr, assuming the Skumanich $P \propto \sqrt t$
      spin-down, while the green, red, and blue curves are $t=1$~Gyr gyrochrones
      \citep{Barnes2010b} for initial periods of 0.12, 1.1, and
      3.4~days, respectively. The gyro-age of NGC~6811 was measured to
      be 1.00$\pm$0.17~Gyr \citep{Janes2013}, which represents the
      oldest calibration of gyrochronology, except for the Sun. Beyond
      that age, there are few measured rotation periods as the known
      open clusters are too far away and faint.
}
    \label{fig1}
   \vspace{-0.4in}
  \end{center}
\end{wrapfigure}

Measuring the age of main-sequence (MS) stars in the Galactic field is, at best, an
imprecise art \citep{Soderblom2010}. Even for solar-type stars, for
which we can tether the scale to the single precise measurement that
exists---that of the Sun---the uncertainties are $\sim$1~Gyr or larger. 
Based on the underlying premise that stars are born with an initial
distribution of rotation periods and slow down over time as they shed
angular momentum \citep{Skumanich1972, Pallavicini1981,
  Soderblom1993b}, \citet{Barnes2003,Barnes2007,Barnes2010b} have
suggested a framework for measuring stellar ages from their rotation
periods. This empirical paradigm, gyrochronology, works well for
solar-type stars in young clusters
\citep[e.g.,][]{Meibom2009,Meibom2011} and, apart from
asteroseismology, is the most precise way to measure stellar
ages of individual stars. Asteroseismology is not practical for a
large number of stars and is impossible for the fainter
ones. Figure~\ref{fig1} shows the results of the 
\textit{Kepler} open cluster study of NGC~6811 that extended the
gyro-age calibrations to $\sim$1~Gyr \citep{Meibom2011a,Janes2013}. 
Beyond that age, there remains a lack of measured rotation periods as the
periods are longer, the photometric variations smaller, and open
clusters further away and therefore fainter. Among the \textit{Kepler} open clusters, it
may be possible to measure rotation periods of MS stars for
the $\sim$2.5~Gyr old NGC~6819 but not for the $\sim$9~Gyr old NGC~6791.

The current gyrochronology age calibration breaks down once the
stars become fully-convective at the spectral type of $\sim$M3--M4. 
In solar-type dwarfs the rotational shear between the convective
and radiative zones is believed to be central in generating
magnetic fields. The spots at the stellar surface are modulated to
the same periods. In fully-convective dwarfs, no such rotational shear
exists. The magnetic fields are presumed to be driven by a turbulent
dynamo \citep[e.g.,][]{Browning2008}. However, they seem to possess
strong magnetic fields of kilogauss scales that give rise to
chromospheric and/or coronal activity seen in the UV, optical, and
X-rays. The active M dwarfs do not always exhibit rotational modulation
\citep{Donati2006,West2009}, perhaps because the convection flows on
the stellar surface are small-scale \citep{Browning2008}. The magnetic
activity in the fully-convective dwarfs lasts much longer, up to
7--8~Gyr for mid--late M dwarfs \citep{West2011}. The relationship
between stellar rotation, magnetic activity, and their evolution with
age is much more complex in fully-convective stars.
The magnetic activity has important ramifications for the habitability
of planets around low-mass stars and serves as a probe of the
internal mechanisms controlling magnetic field generation and stellar
atmospheric heating. In addition, a reliable age determination is key to
understanding the evolution of exoplanet orbits.

The \textit{Kepler} spacecraft, even with only two operational
reaction wheels, is capable of acquiring data to higher precision
($\sim$mmag) and for fainter stars than small- to 
medium-aperture, ground-based telescopes. Moreover, as the stellar
rotation periods are longer and the photometric variations have
smaller amplitudes, the necessary cadence and time baseline cannot be
provided by ground-based instruments. 

\textbf{In this white paper for the Kepler Two-Wheels Program (KTWP),
  we describe how \textit{Kepler}-measured 
  rotation periods of FGKM dwarf binary systems and single M dwarfs
  will enable the calibration of gyrochronology for older dwarfs in
  the Galactic field than currently available and for the fully-convective
  dwarfs at the end of the stellar MS. This will help establish precise
  determinations of stellar ages, to $\sim$10\% for all dwarfs
  that have measurable surface rotations.} The impact of age
measurements in stellar astrophysics as well as in tracing the formation  
history of the Milky Way cannot be overstated. In addition, the
presence of a rich, time-resolved photometry of a large number of
stars will enable many ancillary projects. We briefly describe two
objectives (i) to identify long-period M dwarf eclipsing
binaries to determine the role of magnetic fields and surface rotation
on stellar radii and (ii) to characterize micro-variability among cool
white dwarf (WD) and hot subdwarf B (sdB) stars.

\section{Calibrating the Stellar Rotation--Age Paradigm Using Wide MS Binaries in the \textit{Kepler} Field}
\begin{wrapfigure}{R}{0.6\textwidth}
  \begin{center}
    \vspace{-0.3in}
    \includegraphics[width=0.5\textwidth]{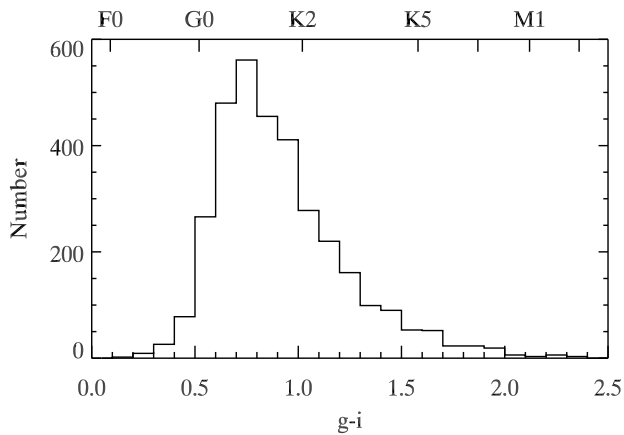}
    \vspace{-0.1in}    \caption{\protect\small
      The $g-i$ color distribution, with the spectral types shown
      along the top x-axis, for the candidate common
      proper motion binaries with at least one component observed by
      \textit{Kepler}. The candidate binaries were identified
      by cross-matching KIC with UCAC4 proper motions
      (K. Weisenburger, \textit{priv. comm.}. Known binaries in the
      Washington Double Star Catalog \citep{Mason2001} are also
      included (K. Janes, T. Mizusawa, \textit{priv. comm.}). The lack
      of M dwarf binaries reflects \textit{Kepler's} selection
      effect and not the dearth of M dwarf binaries in the
      \textit{Kepler} FOV. With a
      wider spread of age and metallicity than available among nearby
      open clusters, these binary systems are central in
      our efforts to benchmark gyrochronology \citep{Barnes2003, Barnes2010a} 
      as well as the chromospheric activity--age relations. With a
      photometric precision and temporal coverage that far exceeds
      what is available from the ground, \textit{Kepler} will be
      critical in measuring rotation periods for these older stars
      where rotation periods can be several months long and photometric
      amplitudes are very small.
}
    \label{fig2}
    \vspace{-0.2in}
  \end{center}
\end{wrapfigure}
Binary (or multiple) star systems are essentially small open clusters, albeit with
just two members. While not able to provide the statistical power of the
dozens of stars typical in an open cluster, an ensemble of binary systems in
the Galactic field can be even more powerful as they span a range of
ages, metallicities, and evolutionary histories that provide a truly
heterogeneous population. Past the age of $\sim$1~Gyr, binaries are also much
closer and brighter than open clusters. As all but the most massive open
clusters, which are further away and fainter, become unbound after
$\sim$1~Gyr, their utility is limited as lower-MS dwarfs tend to
escape first, limiting their availability as age
benchmarks. Therefore, to test and calibrate gyrochronology for older
stars, binary and multiple systems represent powerful keystones.

The \textit{Kepler} FOV contains a large number wide, MS
binary systems. Figure~\ref{fig2} shows the $g-i$ color distribution,
with the inferred spectral types marked along the top x-axis, for $\sim$1600
candidate binary pairs for which at least one component has been
observed by \textit{Kepler} (K. Weisenburger, \textit{priv.
  comm.}). They were identified by matching the \textit{Kepler}
Input Catalog \citep[KIC;][]{Batalha2010} with the UCAC4 proper motion catalog
\citep{Zacharias2013} using the algorithm outlined in \citet{Dhital2010}.
In addition, we found 40 binaries in the Washington Double Star
Catalog \citep{Mason2001} within the \textit{Kepler} FOV (K. Janes,
\textit{priv. comm.}; T. Mizusawa, \textit{priv. comm.}). As
\textit{Kepler} observes only a small fraction of the over 13.2
million stars in the KIC, there are presumably many more binaries that could
be added to the list of stars observed by \textit{Kepler}. The current
observed \textit{Kepler} stars are predominantly FGK dwarfs; thus, the
added targets should be biased towards the mid--late M dwarfs. Given
the baseline of many months and photometric precision of approximately
a mmag, \textit{Kepler} could measure rotation periods for
one or both components of hundreds of binary systems that span the
FGKM spectral types, even accounting for observed binaries that will
not be rotating.

Apart from being the one precise age measurement of a star available to us, the
Sun is also the only age datum we have past $\sim$1~Gyr. There exist
no age measurements that can be used to calibrate gyrochronology, once
rotation periods have been measured. Therefore, to calibrate
the rotation--age relationship, we will have to empirically solve the
three-dimensional plane of rotation period, color (age), and age. In
other words, we will need to identify the tangent field of rotational
isochrones. As this will necessarily be completely empirical, a
solution will require measurements for more than a hundred binaries.
Stellar metallicity is also known to affects magnetic activity levels
\citep[e.g.,][]{Bochanski2011}. If sufficient rotation periods are
measured, we could possibly extend the solution into the fourth
dimension to include metallicity.

Furthermore, a few wide MS$+$WD pairs exist in the KTWP field.  These
evolved pairs offer an opportunity to independently test the
gyrochronology ages obtained from MS stars.  The cooling ages and
masses of the WD components can easily be derived from spectral line
profile fits from ground-based spectroscopy which we are conducting at
NOAO in an ongoing study \citep[e.g.,][]{Zhao2011}.  The WD cooling
age alone provides an absolute lower limit to the companion MS star's
rotationally inferred age.  With masses determined from ground-based
spectral line fits, correction for the MS lifetime of the WD component
can be made, providing an independent check on the total systemic age
derived from rotation and/or activity.  Thus, this subset of wide
pairs in the KTWP field offers a valuable independent consistency
check on the relative ages provided by gyrochronology among MS$+$MS pairs.

\vspace{-0.2in}
\subsection{Comparing the Stellar Rotation--Age and Activity--Age Relations in FGKM Dwarfs}
We propose a rigorous test of two competing approaches
to the measurement of stellar ages: rotation and chromospheric
activity.  These methods often give conflicting results; and there is
much current debate in the scientific community about how best to
determine stellar ages, especially for those not found in the very
limited of accessible clusters \citep{Barnes2009,Soderblom2010}.  In
the \textit{Kepler} field, we have identified a large sample of
wide common proper motion pairs, each consisting of two MS
stars (Figure~\ref{fig2}).  Such pairs provide a very robust basis for determining
rotational age of each pair via gyrochronology enabled by the photometric
micro-variability ($<$1\%) imposed by rotation of spotted stellar
surfaces.  The photometric precision and temporal coverage of the
\textit{Kepler} field, even in its extended mission, far exceeds what is
available from ground-based investigations of the rotation--mass--age
relation.  This is particularly important for pairs older than the Sun
where rotation periods can be several months and photometric
amplitudes are very low.  In parallel to the analysis of the proposed
\textit{Kepler} photometric data, we plan to do follow-up spectroscopic
observations of these pairs to confirm spectral types (mass), to measure
activity proxies such as CaII H and K and {\ha}, and to determine their
metallicity \citep[e.g.,][]{Zhao2011}. The \textit{Kepler} data alone will provide a robust test of
gyrochronology theory and the assumption that such pairs are coeval.
Combined with our ongoing spectroscopy conducted at NOAO, the \textit{Kepler}
data will also support a definitive comparison between
rotation--age and activity--age relations, and how each depends upon
mass and metallicity. 

\vspace{-0.2in}
\subsection{Extending gyrochronology beyond the fully-convective boundary}
M dwarfs are the most numerous stellar constituents of the Milky Way
\citep[$\gtrsim$70\%;][]{Henry1993} and appear to be the most numerous
hosts of terrestrial planets \citep{Howard2012,Dressing2013}. However,
until the recent advent of all-sky surveys, their intrinsic faintness
had long limited our ability to study M dwarfs. As a result, large
discrepancies exist in the empirical (when they exist) and theoretical
values of their masses, radii, and metallicity. Obtaining robust
empirical constraints for these properties, as well as on age, is
critical to characterization of the orbiting, terrestrial planets. 

There are only $\sim$3,000 M dwarfs ($\sim$2\%) among the
observed \textit{Kepler} stars, almost all of which are of early--mid
spectral types \citep{Batalha2010}. Thus, while more than 1,500
rotation periods for M dwarfs were measured from \textit{Kepler} Q3
data \citep{McQuillan2013,Nielsen2013,Reinhold2013},  
only 10--30 were for fully-convective M dwarfs. The vast majority
of measured rotation periods for the $>$M3--M4 dwarfs come from the MEarth survey
\citep{Irwin2011}, with a few from \citet{Kiraga2007}. The period
distribution is bimodal with peaks at $\sim$1~d and $\sim$180~d, with
amplitudes of $\sim$0.01--0.05 mag. The latter peak corresponds to
older stars, possibly associated with the Milky Way's thick disk. The
MEarth survey was able to measure the rotation periods from a
ground-based telescope. With the expected mmag precision, a
higher cadence, and a similar/longer time baseline, KTWP should be
able to measure these rotation periods for the older M dwarfs.
In fact, \citet{McQuillan2013} detected rotation periods for 63\% of
the early--mid M dwarfs in the \textit{Kepler} sample. As the late M
dwarfs are more likely to be active for a longer time, the yield could
be even higher.

\begin{wrapfigure}{R}{0.6\textwidth}
  \begin{center}
    \vspace{-0.3in}
    \includegraphics[width=0.5\textwidth]{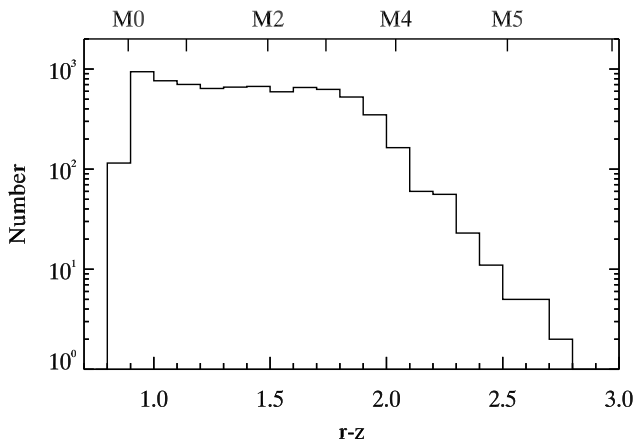}
    \vspace{-0.1in}
    \caption{\protect\small
      The $r-z$ color distribution, with spectral subtypes indicated
      along the top x-axis, for $\sim$7,500 M dwarfs in the
      KIC. This preliminary sample was based on USNO proper
      motions (to select against M giants), which have a known bias
      against the mid--late M dwarfs \citep{Monet2003}. A larger
      sample of M dwarfs has been identified using proper motions from
      the PanStarrs survey (\textit{Kepler} Cycle 5 GO
      proposal; PI: E.~Gaidos), which should be preferentially
      included in KTWP.}
    \label{fig3}
    \vspace{-0.2in}
  \end{center}
\end{wrapfigure}
To include more M dwarfs in KTWP, we conducted a preliminary search
for M dwarfs in KIC. First, we required $r-z >$ 0.89
\citep[$\sim$M0;][]{West2011} and $r <$ 19.0. Then, to select against
M giants, we required $J-H > $ 0.7 \citep{Covey2007} and total proper
motion $>$ 20~{\masyr} \citep[USNO;][]{Monet2003}. Figure~\ref{fig3} shows the 
$r-z$ color distribution for the resultant sample of $\sim$7,500 M
dwarfs. It is highly biased towards the early--mid M dwarfs, with only
$\gtrsim$350 later than M4. This reflects the blue bias of the USNO
proper motions, whose photographic plates were not sensitive enough
for the later M dwarfs. However, a sample of $>$15,000 M dwarfs, that
includes a significant number of mid--late ones, has been identified
from KIC using PanStarrs proper motion (\textit{Kepler} Cycle 5 GO
proposal; PI: E.~Gaidos). There is also an ongoing PanStarrs
project to measure proper motions for stars in the \textit{Kepler}
FOV (PI: N.~Deacon); the preliminary results should be available
well before the KTWP targets are selected (A.~Mann, \textit{priv. comm.})
These mid-late M dwarfs should be preferentially included in KTWP
target lists.


\section{Ancillary Science Goals}
While the focus of this proposal is on measuring the rotation ages of
lower MS stars, two other projects can be undertaken in parallel using
stars in the same KTWP field. 

\subsection{Long-Period Eclipsing Binaries}
Empirical measurements of M dwarf radii are rare \citep{Torres2010}.
Those that have measured radii are either near enough for angular diameter
measurements with optical long-baseline interferometry
\citep[e.g.,][]{Lane2001,Berger2006a,Segransan2003,Boyajian2012},
  or are short-period eclipsing binary stars \citep[e.g.,][]{Stassun2007,Stassun2008,
    Morales2009}.  There currently exist significant 
discrepancies between the radii measured in short-period eclipsing
binary M dwarfs and predictions from evolutionary models \citep{Baraffe1998}.
\citet{Chabrier2007} suggested that M dwarfs in
short-period eclipsing binaries may have inflated radii due to effects
from rapid rotation and magnetic fields; however this hypothesis has
yet to be empirically verified. 

Determining the role of rotation in radius inflation is of
crucial importance to current and future NASA exoplanet missions.
If, in fact, rapidly rotating M dwarfs are inflated compared to slowly
rotating M dwarfs, we must then take this into account when
determining the radius of exoplanets detected by Kepler or NASA's
future TESS mission.  We can test the \citet{Chabrier2007}
hypothesis by searching for and characterizing long-period eclipsing
binary M dwarfs (orbital periods of greater than 10~days), where the
component stars are not synchronously rotating.  \textit{Kepler}
is uniquely suited to this search, as it can observe stars
nearly-continuously and detect the several-hour eclipse signature
within a light curve spanning months to years and can observe
thousands of M dwarfs at once.  Searching for long-period eclipsing
binaries from the ground is extremely challenging with only one
detection to date \citep[LSPM~J1112$+$7626, P $=$ 41~days;][]{Irwin2011a}. 

With eclipse depths of roughly one to several magnitudes, the reduced
photometric performance of KTWP will not inhibit the
detection of long-period eclipsing binary stars.  We propose to search
for long-period eclipsing binary M dwarfs within our proposed M dwarf
sample.  Results will directly feed into the radius determinations of
exoplanets detected by Kepler and TESS. 

\vspace{-0.2in}
\subsection{Variability among Highly Evolved Stars}
A photometric investigation of micro-variability among cool white dwarf
(WD) and hot subdwarf B (sdB) stars is proposed for the extended
Kepler mission.  The Kepler data will enable time series observations
to be obtained for WD targets with absolute magnitude M$_{\textrm V} > +15$ and
color 0.2 $< V-I <$ 1.4 corresponding to the temperature ({\Teff} $<$ 6000~K)
and period ($P <$ 500~s) regime where micro-variability due to
collisionally induced absorption by H$_2$ and/or C$_2$ may occur.  In the
sample of several dozen cool WDs observed so far with ground-based
facilities, no periodic variability larger than $\sim$2\% has been
detected \citep{Rudkin2004}, suggesting such objects have good potential to be used
as faint OIR flux standards for large-aperture and space-based
telescopes.  However, two targets show possible variability near the
limit of what can be confirmed from ground-based photometry.  The
detection of any level of micro-variability (or not) would provide
rigorous tests of cool WD models and the equation of state for
degenerate matter using asteroseismology techniques.

In a parallel project that can be conducted in the
same field, a search for variations in the arrival times of known
pulsations in WD and sdB stars can also be investigated.  Such periodic
changes in the O$-$C diagrams for V391 Pegasi \citep{Silvotti2007} revealed the presence
of a planet in this post-MS system.  Even in its reduced
pointing precision mode, the photometric precision and nearly
continuous multi-year coverage afforded by Kepler will provide several
definitive searches for additional planetary survivors out to several
astronomical units from the host stars.

\section{Observing Strategy with \textit{Kepler}'s Two-Wheel Program}
As a result of only two reaction wheels being in working condition,
KTWP will be able to deliver photometric precision of 0.3--0.6 mmag
rms to 1 mmag rms depending on the seeing. Even accounting for pixel
sensitivity variations, the relative photometry is expected to be
$\sim$0.3--1\%\footnote{\url{http://keplergo.arc.nasa.gov/News.shtml\#TwoWheelWhitePaper}}
for all stars brighter than $\sim$16--17~mag. Photometric variations
of FGKM dwarfs, modulated to stellar surface rotation, are typically
0.01--0.05~mag \citep{McQuillan2013,Nielsen2013,Reinhold2013}. The
rotation periods for dwarfs up to 16--17~mag will 
be easily detected. In fact, under the best performance scenarios, it might
even be possible to detect giant planets. For fainter dwarfs,
detection efficiency will be dependent on the precision ultimately
delivered by KTWP and the amplitude of their photometric variations.

We have constructed preliminary samples of $\gtrsim$1600 wide FGKM
dwarf binaries and $\gtrsim$7,500 single M dwarfs in the current
\textit{Kepler} FOV. There are a handful
of wide WD$+$MS systems, single WDs, and sdBs as well. While our sample
has a distinct lack of mid--late M dwarfs due to its dependence on
USNO proper motions, such a sample has recently been constructed
(\textit{Kepler} Cycle 5 GO proposal; PI: E~Gaidos). To
cover a wide range of parameter space that includes mass, age,
activity level, and metallicity, we request a sample of 1000 binary systems
and 2000 single M dwarfs be observed. However, the number of targets
can certainly be changed to maximize \textit{Kepler}'s utility.
All of these targets can be observed with the long cadence (30~min
samples) used in the regular \textit{Kepler} operations. Our proposed
programs will not be affected by the need to reorient the spacecraft
every four days.

This white paper is predicated upon \textit{Kepler} pointing at the
current FOV, which is reasonable given the time and effort already 
spent in characterizing the stars in this FOV. This was done under the
assumption that pointing elsewhere will require either a significant
effort in identifying $\sim$150,000 targets that \textit{Kepler} is
capable of monitoring at a time or a dramatic under-utilization of
\textit{Kepler}'s capabilities. \textbf{However, we would like to note that
our science goals are not dependent on where \textit{Kepler} is
pointed.  There exist large samples of both FGKM binaries
\citep{Luyten1997,Chaname2004,Lepine2007a,Dhital2010} and
fully-convective dwarfs \citep{Bochanski2010,West2011} throughout the
sky.} Some parts of the sky, particularly in the Sloan Digital Sky
Survey \citep{York2000} footprint, might even be better due to the
extant multi-band data.

\section{Conclusion}
This white paper outlines several projects which can easily be
undertaken with the KTWP within the reduced photometric precision
available.  None of these projects will require any significant
movement of the spacecraft from its present field.

First, we propose the first comprehensive calibration of the
rotation--age--mass--metallicity relation to the bottom of the MS and to
nearly the age of the Galaxy’s disk ($\sim$10~Gyr).  We have determined
that the KTWP field contains hundreds of lower-MS stars that are in
widely separated non-interacting common proper motion pairs.
Components of a given pair are expected to be of the same age and
chemical composition.  But are they?  The rotational ages derived from
the proposed KTWP will provide the first direct test of the notion
that such pairs are coeval.  Moreover, the range of spectral types
(i.e., masses) and metallicities represented by these pairs will permit
the first comprehensive examination of the
rotation--age--mass--metallicity relation among MS stars of M spectral
type across the transition regime where self-generating toroidal
fields, like the Sun’s, give way to turbulent dynamo-driven magnetic
fields.  In conjunction with ground-based spectroscopy of
chromospheric activity among these wide pairs which is already in
progress at NOAO facilities, the proposed KTWP observations will allow
a detailed comparison of the consistency and precision of ages derived
from activity vs. rotation as a function of spectral type (mass).
Moreover, combination of the KTWP photometry and ground-based
spectroscopy will permit the inclusion of a fourth dimension
(metallicity) in the analysis.  Finally, a few wide pairs contain a MS
paired with a WD.  In such cases, the cooling age of the latter
provides a firm lower limit to the total age of the system, hence a
valuable limit on both the rotational and activity ages.

Our second proposed project involves eclipsing binaries containing
late-type MS stars in the KTWP field. Empirical measurements of M
dwarf radii are rare, with most coming from short-period eclipsing
binaries. However, significant discrepancies ($\gtrsim$10\%) 
exist between the measured radii and the predictions from evolutionary
models. As the short-period binaries may have inflated radii,
long-period eclipsing binaries are necessary to test the evolutionary
models. With similar brightness components, eclipse depths are roughly
one to several magnitudes. A large database or time-resolved
photometry would be a trove for eclipsing binary searches.

Our third proposed project involves a search for microvariability
among evolved stars in the KTWP field.  Several WDs and sdB stars are
already known to be in the Kepler field.  Those already known to
exhibit non-radial pulsations will be monitored during the extended
time afforded by the KTWP to search for light time modulation of the
arriving pulses, the signature of planets or substellar companions.
In addition, a few cool WDs in the field may have atmospheric
opacities dominated by dissociation of H$_2$ or C$_2$ molecules that could
drive pulsations.  The photometric precision afforded by the KTWP
should provide a conclusive test of this hypothesis.

While we hope to conduct all of the above projects in parallel, any
one of them can easily be accommodated separately within the
constraints of the KTWP scenario.  Some modes overlap exists between
the targets already observed over the past several years with Kepler
and our proposed sample.  In these cases we will be able to look for
the signatures of activity/spot cycles.  In cases where the
photometric modulation has extended over many rotational or
pulsational cycles, we will be able to search for the signature of
light-time modulation in the O$-$C diagrams caused by planets or
substellar companions.

\begin{multicols}{2}
\bibliographystyle{apj}
\bibliography{ads}

\begin{thebibliography}{51}
\expandafter\ifx\csname natexlab\endcsname\relax\def\natexlab#1{#1}\fi

\bibitem[{{Baraffe} {et~al.}(1998){Baraffe}, {Chabrier}, {Allard}, \&
  {Hauschildt}}]{Baraffe1998}
{Baraffe}, I., {Chabrier}, G., {Allard}, F., \& {Hauschildt}, P.~H. 1998, \aap,
  337, 403

\bibitem[{{Barnes}(2003)}]{Barnes2003}
{Barnes}, S.~A. 2003, \apj, 586, 464

\bibitem[{{Barnes}(2007)}]{Barnes2007}
---. 2007, \apj, 669, 1167

\bibitem[{{Barnes}(2009)}]{Barnes2009}
{Barnes}, S.~A. 2009, in IAU Symposium, Vol. 258, IAU Symposium, ed. E.~E.
  {Mamajek}, D.~R. {Soderblom}, \& R.~F.~G. {Wyse}, 345--356

\bibitem[{{Barnes}(2010)}]{Barnes2010b}
---. 2010, \apj, 721

\bibitem[{{Barnes} \& {Kim}(2010)}]{Barnes2010a}
{Barnes}, S.~A., \& {Kim}, Y. 2010, \apj, 721, 675

\bibitem[{{Batalha} {et~al.}(2010){Batalha}, {Borucki}, {Koch}, {Bryson},
  {Haas}, {Brown}, {Caldwell}, {Hall}, {Gilliland}, {Latham}, {Meibom}, \&
  {Monet}}]{Batalha2010}
{Batalha}, N.~M., {Borucki}, W.~J., {Koch}, D.~G., {et~al.} 2010, \apjl, 713,
  L109

\bibitem[{{Berger} {et~al.}(2006){Berger}, {Gies}, {McAlister}, {ten
  Brummelaar}, {Henry}, {Sturmann}, {Sturmann}, {Turner}, {Ridgway},
  {Aufdenberg}, \& {M{\'e}rand}}]{Berger2006a}
{Berger}, D.~H., {Gies}, D.~R., {McAlister}, H.~A., {et~al.} 2006, \apj, 644,
  475

\bibitem[{{Bochanski} {et~al.}(2010){Bochanski}, {Hawley}, {Covey}, {West},
  {Reid}, {Golimowski}, \& {Ivezi{\'c}}}]{Bochanski2010}
{Bochanski}, J.~J., {Hawley}, S.~L., {Covey}, K.~R., {et~al.} 2010, \aj, 139,
  2679

\bibitem[{{Bochanski} {et~al.}(2011){Bochanski}, {Hawley}, \&
  {West}}]{Bochanski2011}
{Bochanski}, J.~J., {Hawley}, S.~L., \& {West}, A.~A. 2011, \aj, 141, 98

\bibitem[{{Boyajian} {et~al.}(2012){Boyajian}, {McAlister}, {van Belle},
  {Gies}, {ten Brummelaar}, {von Braun}, {Farrington}, {Goldfinger}, {O'Brien},
  {Parks}, {Richardson}, {Ridgway}, {Schaefer}, {Sturmann}, {Sturmann},
  {Touhami}, {Turner}, \& {White}}]{Boyajian2012}
{Boyajian}, T.~S., {McAlister}, H.~A., {van Belle}, G., {et~al.} 2012, \apj,
  746, 101

\bibitem[{{Browning}(2008)}]{Browning2008}
{Browning}, M.~K. 2008, \apj, 676, 1262

\bibitem[{{Chabrier} {et~al.}(2007){Chabrier}, {Gallardo}, \&
  {Baraffe}}]{Chabrier2007}
{Chabrier}, G., {Gallardo}, J., \& {Baraffe}, I. 2007, \aap, 472, L17

\bibitem[{{Chanam{\'e}} \& {Gould}(2004)}]{Chaname2004}
{Chanam{\'e}}, J., \& {Gould}, A. 2004, \apj, 601, 289

\bibitem[{{Covey} {et~al.}(2007){Covey}, {Ivezi{\'c}}, {Schlegel},
  {Finkbeiner}, {Padmanabhan}, {Lupton}, {Ag{\"u}eros}, {Bochanski}, {Hawley},
  {West}, {Seth}, {Kimball}, {Gogarten}, {Claire}, {Haggard}, {Kaib},
  {Schneider}, \& {Sesar}}]{Covey2007}
{Covey}, K.~R., {Ivezi{\'c}}, {\v Z}., {Schlegel}, D., {et~al.} 2007, \aj, 134,
  2398

\bibitem[{{Dhital} {et~al.}(2010){Dhital}, {West}, {Stassun}, \&
  {Bochanski}}]{Dhital2010}
{Dhital}, S., {West}, A.~A., {Stassun}, K.~G., \& {Bochanski}, J.~J. 2010, \aj,
  139, 2566

\bibitem[{{Donati} {et~al.}(2006){Donati}, {Forveille}, {Collier Cameron},
  {Barnes}, {Delfosse}, {Jardine}, \& {Valenti}}]{Donati2006}
{Donati}, J.-F., {Forveille}, T., {Collier Cameron}, A., {et~al.} 2006,
  Science, 311, 633

\bibitem[{{Dressing} \& {Charbonneau}(2013)}]{Dressing2013}
{Dressing}, C.~D., \& {Charbonneau}, D. 2013, \apj, 767, 95

\bibitem[{{Henry} \& {McCarthy}(1993)}]{Henry1993}
{Henry}, T.~J., \& {McCarthy}, D.~W.~J. 1993, \aj, 106, 773

\bibitem[{{Howard} {et~al.}(2012){Howard}, {Marcy}, {Bryson}, {Jenkins},
  {Rowe}, {Batalha}, {Borucki}, {Koch}, {Dunham}, {Gautier}, {Van Cleve},
  {Cochran}, {Latham}, {Lissauer}, {Torres}, {Brown}, {Gilliland}, {Buchhave},
  {Caldwell}, {Christensen-Dalsgaard}, {Ciardi}, {Fressin}, {Haas}, {Howell},
  {Kjeldsen}, {Seager}, {Rogers}, {Sasselov}, {Steffen}, {Basri},
  {Charbonneau}, {Christiansen}, {Clarke}, {Dupree}, {Fabrycky}, {Fischer},
  {Ford}, {Fortney}, {Tarter}, {Girouard}, {Holman}, {Johnson}, {Klaus},
  {Machalek}, {Moorhead}, {Morehead}, {Ragozzine}, {Tenenbaum}, {Twicken},
  {Quinn}, {Isaacson}, {Shporer}, {Lucas}, {Walkowicz}, {Welsh}, {Boss},
  {Devore}, {Gould}, {Smith}, {Morris}, {Prsa}, {Morton}, {Still}, {Thompson},
  {Mullally}, {Endl}, \& {MacQueen}}]{Howard2012}
{Howard}, A.~W., {Marcy}, G.~W., {Bryson}, S.~T., {et~al.} 2012, \apjs, 201, 15

\bibitem[{{Irwin} {et~al.}(2011{\natexlab{a}}){Irwin}, {Berta}, {Burke},
  {Charbonneau}, {Nutzman}, {West}, \& {Falco}}]{Irwin2011}
{Irwin}, J., {Berta}, Z.~K., {Burke}, C.~J., {et~al.} 2011{\natexlab{a}}, \apj,
  727, 56

\bibitem[{{Irwin} {et~al.}(2011{\natexlab{b}}){Irwin}, {Quinn}, {Berta},
  {Latham}, {Torres}, {Burke}, {Charbonneau}, {Dittmann}, {Esquerdo},
  {Stefanik}, {Oksanen}, {Buchhave}, {Nutzman}, {Berlind}, {Calkins}, \&
  {Falco}}]{Irwin2011a}
{Irwin}, J.~M., {Quinn}, S.~N., {Berta}, Z.~K., {et~al.} 2011{\natexlab{b}},
  \apj, 742, 123

\bibitem[{{Janes} {et~al.}(2013){Janes}, {Barnes}, {Meibom}, \&
  {Hoq}}]{Janes2013}
{Janes}, K., {Barnes}, S.~A., {Meibom}, S., \& {Hoq}, S. 2013, \aj, 145, 7

\bibitem[{{Kiraga} \& {Stepien}(2007)}]{Kiraga2007}
{Kiraga}, M., \& {Stepien}, K. 2007, Acta Astronomica, 57, 149

\bibitem[{{Lane} {et~al.}(2001){Lane}, {Boden}, \& {Kulkarni}}]{Lane2001}
{Lane}, B.~F., {Boden}, A.~F., \& {Kulkarni}, S.~R. 2001, \apjl, 551, L81

\bibitem[{{L{\'e}pine} \& {Bongiorno}(2007)}]{Lepine2007a}
{L{\'e}pine}, S., \& {Bongiorno}, B. 2007, \aj, 133, 889

\bibitem[{{Luyten}(1997)}]{Luyten1997}
{Luyten}, W.~J. 1997, VizieR Online Data Catalog, 1130, 0

\bibitem[{{Mason} {et~al.}(2001){Mason}, {Wycoff}, {Hartkopf}, {Douglass}, \&
  {Worley}}]{Mason2001}
{Mason}, B.~D., {Wycoff}, G.~L., {Hartkopf}, W.~I., {Douglass}, G.~G., \&
  {Worley}, C.~E. 2001, \aj, 122, 3466

\bibitem[{{McQuillan} {et~al.}(2013){McQuillan}, {Aigrain}, \&
  {Mazeh}}]{McQuillan2013}
{McQuillan}, A., {Aigrain}, S., \& {Mazeh}, T. 2013, \mnras, 432, 1203

\bibitem[{{Meibom} {et~al.}(2009){Meibom}, {Mathieu}, \&
  {Stassun}}]{Meibom2009}
{Meibom}, S., {Mathieu}, R.~D., \& {Stassun}, K.~G. 2009, \apj, 695, 679

\bibitem[{{Meibom} {et~al.}(2011{\natexlab{a}}){Meibom}, {Mathieu}, {Stassun},
  {Liebesny}, \& {Saar}}]{Meibom2011}
{Meibom}, S., {Mathieu}, R.~D., {Stassun}, K.~G., {Liebesny}, P., \& {Saar},
  S.~H. 2011{\natexlab{a}}, \apj, 733, 115

\bibitem[{{Meibom} {et~al.}(2011{\natexlab{b}}){Meibom}, {Barnes}, {Latham},
  {Batalha}, {Borucki}, {Koch}, {Basri}, {Walkowicz}, {Janes}, {Jenkins}, {Van
  Cleve}, {Haas}, {Bryson}, {Dupree}, {Furesz}, {Szentgyorgyi}, {Buchhave},
  {Clarke}, {Twicken}, \& {Quintana}}]{Meibom2011a}
{Meibom}, S., {Barnes}, S.~A., {Latham}, D.~W., {et~al.} 2011{\natexlab{b}},
  \apjl, 733, L9

\bibitem[{{Monet} {et~al.}(2003){Monet}, {Levine}, {Canzian}, {Ables}, {Bird},
  {Dahn}, {Guetter}, {Harris}, {Henden}, {Leggett}, {Levison}, {Luginbuhl},
  {Martini}, {Monet}, {Munn}, {Pier}, {Rhodes}, {Riepe}, {Sell}, {Stone},
  {Vrba}, {Walker}, {Westerhout}, {Brucato}, {Reid}, {Schoening}, {Hartley},
  {Read}, \& {Tritton}}]{Monet2003}
{Monet}, D.~G., {Levine}, S.~E., {Canzian}, B., {et~al.} 2003, \aj, 125, 984

\bibitem[{{Morales} {et~al.}(2009){Morales}, {Ribas}, {Jordi}, {Torres},
  {Gallardo}, {Guinan}, {Charbonneau}, {Wolf}, {Latham}, {Anglada-Escud{\'e}},
  {Bradstreet}, {Everett}, {O'Donovan}, {Mandushev}, \&
  {Mathieu}}]{Morales2009}
{Morales}, J.~C., {Ribas}, I., {Jordi}, C., {et~al.} 2009, \apj, 691, 1400

\bibitem[{{Nielsen} {et~al.}(2013){Nielsen}, {Gizon}, {Schunker}, \&
  {Karoff}}]{Nielsen2013}
{Nielsen}, M.~B., {Gizon}, L., {Schunker}, H., \& {Karoff}, C. 2013, \aap, 557,
  L10

\bibitem[{{Pallavicini} {et~al.}(1981){Pallavicini}, {Golub}, {Rosner},
  {Vaiana}, {Ayres}, \& {Linsky}}]{Pallavicini1981}
{Pallavicini}, R., {Golub}, L., {Rosner}, R., {et~al.} 1981, \apj, 248, 279

\bibitem[{{Reinhold} {et~al.}(2013){Reinhold}, {Reiners}, \&
  {Basri}}]{Reinhold2013}
{Reinhold}, T., {Reiners}, A., \& {Basri}, G. 2013, ArXiv e-prints

\bibitem[{{Rudkin}(2004)}]{Rudkin2004}
{Rudkin}, M. 2004, M.~S. Thesis, Florida Institute of Technology

\bibitem[{{S{\'e}gransan} {et~al.}(2003){S{\'e}gransan}, {Kervella},
  {Forveille}, \& {Queloz}}]{Segransan2003}
{S{\'e}gransan}, D., {Kervella}, P., {Forveille}, T., \& {Queloz}, D. 2003,
  \aap, 397, L5

\bibitem[{{Silvotti} {et~al.}(2007){Silvotti}, {Schuh}, {Janulis}, {Solheim},
  {Bernabei}, {{\O}stensen}, {Oswalt}, {Bruni}, {Gualandi}, {Bonanno},
  {Vauclair}, {Reed}, {Chen}, {Leibowitz}, {Paparo}, {Baran}, {Charpinet},
  {Dolez}, {Kawaler}, {Kurtz}, {Moskalik}, {Riddle}, \& {Zola}}]{Silvotti2007}
{Silvotti}, R., {Schuh}, S., {Janulis}, R., {et~al.} 2007, \nat, 449, 189

\bibitem[{{Skumanich}(1972)}]{Skumanich1972}
{Skumanich}, A. 1972, \apj, 171, 565

\bibitem[{{Soderblom}(2010)}]{Soderblom2010}
{Soderblom}, D.~R. 2010, \araa, 48, 581

\bibitem[{{Soderblom} {et~al.}(1993){Soderblom}, {Stauffer}, {MacGregor}, \&
  {Jones}}]{Soderblom1993b}
{Soderblom}, D.~R., {Stauffer}, J.~R., {MacGregor}, K.~B., \& {Jones}, B.~F.
  1993, \apj, 409, 624

\bibitem[{{Stassun} {et~al.}(2008){Stassun}, {Mathieu}, {Cargile}, {Aarnio},
  {Stempels}, \& {Geller}}]{Stassun2008}
{Stassun}, K.~G., {Mathieu}, R.~D., {Cargile}, P.~A., {et~al.} 2008, \nat, 453,
  1079

\bibitem[{{Stassun} {et~al.}(2007){Stassun}, {Mathieu}, \&
  {Valenti}}]{Stassun2007}
{Stassun}, K.~G., {Mathieu}, R.~D., \& {Valenti}, J.~A. 2007, \apj, 664, 1154

\bibitem[{{Torres} {et~al.}(2010){Torres}, {Andersen}, \&
  {Gim{\'e}nez}}]{Torres2010}
{Torres}, G., {Andersen}, J., \& {Gim{\'e}nez}, A. 2010, \aapr, 18, 67

\bibitem[{{West} \& {Basri}(2009)}]{West2009}
{West}, A.~A., \& {Basri}, G. 2009, \apj, 693, 1283

\bibitem[{{West} {et~al.}(2011){West}, {Morgan}, {Bochanski}, {Andersen},
  {Bell}, {Kowalski}, {Davenport}, {Hawley}, {Schmidt}, {Bernat}, {Hilton},
  {Muirhead}, {Covey}, {Rojas-Ayala}, {Schlawin}, {Gooding}, {Schluns},
  {Dhital}, {Pineda}, \& {Jones}}]{West2011}
{West}, A.~A., {Morgan}, D.~P., {Bochanski}, J.~J., {et~al.} 2011, \aj, 141, 97

\bibitem[{{York} {et~al.}(2000){York}, {Adelman}, {Anderson}, {Anderson},
  {Annis}, {Bahcall}, {Bakken}, {Barkhouser}, {Bastian}, {Berman}, {Boroski},
  {Bracker}, {Briegel}, {Briggs}, {Brinkmann}, {Brunner}, {Burles}, {Carey},
  {Carr}, {Castander}, {Chen}, {Colestock}, {Connolly}, {Crocker}, {Csabai},
  {Czarapata}, {Davis}, {Doi}, {Dombeck}, {Eisenstein}, {Ellman}, {Elms},
  {Evans}, {Fan}, {Federwitz}, {Fiscelli}, {Friedman}, {Frieman}, {Fukugita},
  {Gillespie}, {Gunn}, {Gurbani}, {de Haas}, {Haldeman}, {Harris}, {Hayes},
  {Heckman}, {Hennessy}, {Hindsley}, {Holm}, {Holmgren}, {Huang}, {Hull},
  {Husby}, {Ichikawa}, {Ichikawa}, {Ivezi{\'c}}, {Kent}, {Kim}, {Kinney},
  {Klaene}, {Kleinman}, {Kleinman}, {Knapp}, {Korienek}, {Kron}, {Kunszt},
  {Lamb}, {Lee}, {Leger}, {Limmongkol}, {Lindenmeyer}, {Long}, {Loomis},
  {Loveday}, {Lucinio}, {Lupton}, {MacKinnon}, {Mannery}, {Mantsch}, {Margon},
  {McGehee}, {McKay}, {Meiksin}, {Merelli}, {Monet}, {Munn}, {Narayanan},
  {Nash}, {Neilsen}, {Neswold}, {Newberg}, {Nichol}, {Nicinski}, {Nonino},
  {Okada}, {Okamura}, {Ostriker}, {Owen}, {Pauls}, {Peoples}, {Peterson},
  {Petravick}, {Pier}, {Pope}, {Pordes}, {Prosapio}, {Rechenmacher}, {Quinn},
  {Richards}, {Richmond}, {Rivetta}, {Rockosi}, {Ruthmansdorfer}, {Sandford},
  {Schlegel}, {Schneider}, {Sekiguchi}, {Sergey}, {Shimasaku}, {Siegmund},
  {Smee}, {Smith}, {Snedden}, {Stone}, {Stoughton}, {Strauss}, {Stubbs},
  {SubbaRao}, {Szalay}, {Szapudi}, {Szokoly}, {Thakar}, {Tremonti}, {Tucker},
  {Uomoto}, {Vanden Berk}, {Vogeley}, {Waddell}, {Wang}, {Watanabe},
  {Weinberg}, {Yanny}, {Yasuda}, \& {SDSS Collaboration}}]{York2000}
{York}, D.~G., {Adelman}, J., {Anderson}, Jr., J.~E., {et~al.} 2000, \aj, 120,
  1579

\bibitem[{{Zacharias} {et~al.}(2013){Zacharias}, {Finch}, {Girard}, {Henden},
  {Bartlett}, {Monet}, \& {Zacharias}}]{Zacharias2013}
{Zacharias}, N., {Finch}, C.~T., {Girard}, T.~M., {et~al.} 2013, \aj, 145, 44

\bibitem[{{Zhao} {et~al.}(2011){Zhao}, {Oswalt}, {Rudkin}, {Zhao}, \&
  {Chen}}]{Zhao2011}
{Zhao}, J.~K., {Oswalt}, T.~D., {Rudkin}, M., {Zhao}, G., \& {Chen}, Y.~Q.
  2011, \aj, 141, 107

\end{thebibliography}
\end{multicols}

\end{document}